\documentclass{article}

\usepackage{PRIMEarxiv}

\usepackage[utf8]{inputenc} 
\usepackage[T1]{fontenc}    
\usepackage{hyperref}       
\usepackage{url}            
\usepackage{booktabs}       
\usepackage{amsfonts}       
\usepackage{nicefrac}       
\usepackage{microtype}      
\usepackage{lipsum}
\usepackage{fancyhdr}       
\usepackage{graphicx}       
\graphicspath{{media/}}     
\usepackage{cite}
\usepackage{amsmath,amssymb,amsfonts}
\usepackage{algorithmic}
\usepackage{textcomp}
\usepackage{multirow}
\usepackage{setspace}
\title{Mental Workload Estimation with Electroencephalogram Signals by Combining Multi-Space Deep Models
}

\author{
  Hong-Hai Nguyen, Ngumimi Karen Iyortsuun, Seungwon Kim, Hyung-Jeong Yang, and Soo-Hyung Kim \thanks{\textit{Corresponding author: Soo-Hyung Kim}} \\
  Artificial Intelligence Convergence, Chonnam National University,
Gwangju, 61186, South Korea \\ honghaik14@gmail.com;
kareniyortsuun@gmail.com; Seungwon.Kim@jnu.ac.kr; hjyang@jnu.ac.kr; shkim@jnu.ac.kr.
}

\begin{document}
\maketitle

\begin{abstract}
The human brain remains continuously active, whether an individual is working or at rest. Mental activity is a daily process, and if the brain becomes excessively active, known as overload, it can adversely affect human health. Recently, advancements in early prediction of mental health conditions have emerged, aiming to prevent serious consequences and enhance the overall quality of life. Consequently, the estimation of mental status has garnered significant attention from diverse researchers due to its potential benefits. While various signals are employed to assess mental state, the electroencephalogram, containing extensive information about the brain, is widely utilized by researchers. In this paper, we categorize mental workload into three states (low, middle, and high) and estimate a continuum of mental workload levels. Our method leverages information from multiple spatial dimensions to achieve optimal results in mental estimation. For the time domain approach, we employ Temporal Convolutional Networks. In the frequency domain, we introduce a novel architecture based on combining residual blocks, termed the Multi-Dimensional Residual Block. The integration of these two domains yields significant results compared to individual estimates in each domain. Our approach achieved a 74.98\% accuracy in the three-class classification, surpassing the provided data results at 69.00\%. Specially, our method demonstrates efficacy in estimating continuous levels, evidenced by a corresponding Concordance Correlation Coefficient ($CCC$) result of 0.629. The combination of time and frequency domain analysis in our approach highlights the exciting potential to improve healthcare applications in the future.
\end{abstract}


\section{Introduction}\label{sec1}
Mental workload (MWL) refers to brain activities, which are the number of resources in the human brain. The level of the resource can be changed when the human is thinking or performing a task ~\cite{moray2013mental}. Because the mental workload resource of a person is limited, the ability to process a lot of information at once will be limited \cite{gomez2021studying, kirschner2010minimal}. Humans will be bored if they do easy work when their MWL is low and their MWL is high when they do complicated tasks. However, a high MWL is not suitable for their health \cite{zhang2019spectral} and can impair their abilities for memory, communication, activity, etc. For jobs with high MWL, such as doctors, soldiers and pilots, the brain sometimes has serious accidents \cite{zhang2018learning}. Furthermore, understanding how the human brain works in daily activities and tasks is an essential area of neuroergonomics research. Therefore, the MWL estimation helps to observe and help people at work and evaluate their work system, which can be improved in the future \cite{wang2015using}. Because the EEG signal is a recording of the biological activities of individual brain cells or a subset of brain cells transmitted directly or indirectly through the cerebral cortex and scalp. EEG signals that record brain activities reflect physiological and pathological functions of the hemisphere or of the whole brain related to clinical symptoms, supplementing diagnosis and monitoring treatment, called a clinical electroencephalogram. By providing a reliable EEG can contribute to predicting and preventing work-related risks. Diagnostic applications in MWL estimation can be created to address various real-world issues in healthcare, education, and smart traffic, among others \cite{dimitrakopoulos2017task,bannert2002managing}.

Three metrics were used to measure MWL, namely subject, performance, and physiological measures \cite{wang2015using, chakladar2020eeg, gwizdka2010distribution, cegarra2008use}. The traditional method to measure the context is through a set of various questions, such as the Aeronautics and Space Administration Task Load Index (NASA-TLX) \cite{hart1988development} or Assessment Technique (SWAT) \cite{reid1988subjective}. Performance measurement involves measuring the person's performance during the task with an increasing workload. Both subject and performance measures are taken after the task is completed, making them prone to bias. On the contrary, physiological measurements can continuously record information about the workload and do not affect the performance of the main task. Therefore, physiological signals can be used to evaluate the effectiveness of mental workload.

Physiological measurement uses a variety of biosignals such as electroencephalogram (EEG), Galvanic Skin Response (GSR), Heart Rate Variability (HRV), Skin Temperature (Temp), R-R intervals (RR) and electrocardiogram (ECG). Among such signals, EEG is widely used for the estimation of MWL \cite{taori2022cognitive} because the information capacity in EEG is higher than other physiological signals \cite{hogervorst2014combining}. However, most researchers do not release their datasets, which makes it challenging to compare methods. Fortunately, Lim et al. \cite{lim2018stew} provided an open dataset with a large number of participants and MWL levels. In this study, we conducted experiments on this dataset to estimate the MWL. Our work comprises classification (low, middle and high) and continuous levels estimation (our proposed) for MWL. Specifically, we proposed a Multi-Dimensional Residual Block to learn frequency domain information, and we combined time and frequency domain information based on multimodal fusion, which significantly improved the results.


In conclusion, our contributions are summarized as follows:
\begin{itemize}
    \item Proposed Multi-Dimensional Residual Block to learn information about the frequency domain.
    \item Combined time and frequency domains based on multimodal fusion. 
Specifically, we conducted various experiments to demonstrate the combination method that proved to be efficient.
    \item Solved to classification and continuous levels estimation for mental workload. In this study, we proposed a method for estimating the continuum of cognitive estimation, a task not addressed by previous methods.
\end{itemize}

The next parts of our paper are presented in the following sections: related work is introduced in \autoref{sec:related_work}, the proposed method is described in \autoref{sec:proposed_method}, the experiments in \autoref{sec:experiments} and the discussion in \autoref{sec:discussion} and finally, our work concludes in \autoref{sec:conclusion}. 

\section{Related work}\label{sec:related_work}
Traditional Machine learning and Deep learning are widely used methods to classify MWL. Handcrafted features usually combine with traditional methods, while deep learning automatically extracts features for classification. Wang et al. \cite{wang2015using} employed Support Vector Machine (SVM) to classify different levels of MWL, including low workload (0-back) and high workload (1, 2, 3-back), as well as distinguishing between different levels of workload such as 1-back and 2-back (80\%) and 1-back and 3-back (84\%). The authors extracted a range of features based on statistical, signal power, morphological, and frequency features for binary classification. Ladekar et al. \cite{ladekar2021eeg} utilized K-nearest neighbors (KNN) to classify four levels of cognitive load by decomposing the EEG signal into ten subbands and combining it with ensemble subspace KNN. Their method achieved an accuracy of 83.65\% for four classes. However, the authors performed the experiment on a small sample size of less than 2000, limiting the generalizability of their findings. Gómez et al. \cite{gomez2021studying} used SVM, KNN, and random forest (RF), which are combined with features extraction from frequency bands, the energy and the entropy of the Discrete Wavelet Transform (DWT), Hjorth parameters, Fractal dimension, Detrended fluctuation analysis, and Lempel-Ziv complexity. Zammouri et al. \cite{zammouri2018brain} used PSD and archived an accuracy of $79\%$ in the theta band. The alpha band got $78\%$ of the average accuracy. There are two levels of cognitive load used to classify. The authors showed that theta and alpha powers decrease with the increasing level of mental task in the central and posterior locations, respectively. The research methods employ the Machine Learning technique for classification. To achieve desirable outcomes, it necessitates the calculation and appropriate selection of features. However, the drawback of these methods is that they do not incorporate continuous information of the time or frequency domain, which rely on handcrafted features.

Automatic feature extraction of the deep learning model is better than handcrafted features. In their work, Yang et al. \cite{yang2019assessing} achieved a classification accuracy of $92\%$ for MWL by using SDAE (Stacked Denoising Autoencoder) to preserve local information in the EEG. They also applied an ensemble classifier to classify binary MWL (low and high) by extracting PSD features (theta, alpha, beta, and gamma). Qiao et al. \cite{qiao2020ternary} proposed a Ternary Task Convolutional Bidirectional Neural Turing Machine (TT-CBNTM) for the classification of cognitive states, which are four levels of cognitive load. The TT-CBNTM learns temporal information and preserves spectral and spatial information of EEG. Moreover, the TT-CBNTM can effectively avoid overfitting. Although the method gives quite high results (96.3\%), the authors performed it on a small sample (2600 samples). Kwak et al. \cite{kwak2020multilevel} achieved a binary classification accuracy of $93.9\%$ using a combination of a three-dimensional convolutional neural network (3D CNN) and a long short-term memory model (LSTM). They obtained 3D EEG images from the 1D EEG signal, and the 3D-CNN learned spatial and spectral information from these images, while the LSTM extracted temporal attention from subframes of the EEG image. Deep learning techniques have the capability to extract features automatically. However, previous studies have predominantly used frequency domain information as input for the models. As a result, the outcomes indicate that the potential of time-domain information has not been fully utilized.

In the field of emotion and motor imagery, a 3D representation of the EEG signal has been used. To maintain spatial information on the electrodes, Zhao et al. \cite{zhao2019multi} converted the 2D array of EEG signals into 3D. They applied a multi-branch 3D-CNN to classify motor imagery using the 3D representation, which can be extended to other related areas such as cognitive load and seizure detection. Similarly, Liu et al. \cite{liu2021three} utilized 3D representation to classify motor stages while preserving spatial and temporal information, by passing the 3D representation through a three-branch 3D-CNN. Jia et al. \cite{jia2020sst} used two 3D representations (spectral feature and temporal feature) to complement the significant difference between features. For EEG emotion recognition, a spatial-spectral-temporal based attention 3D dense network was proposed, which performed well on two datasets.

We propose a fusion of the time and frequency domains to leverage the richness of information available in both spaces for the classification task. It is noteworthy that most existing methods are focused on classification, and a regression approach for the continuous estimation of MWL remains relatively less explored. In this study, we adopt a regression framework for continuous MWL estimation. Furthermore, we exploit the benefits of 3D representation to facilitate the learning of complex relationships among the features extracted from EEG signals. 

\begin{figure*}
\centerline{\includegraphics[width=35.1pc]{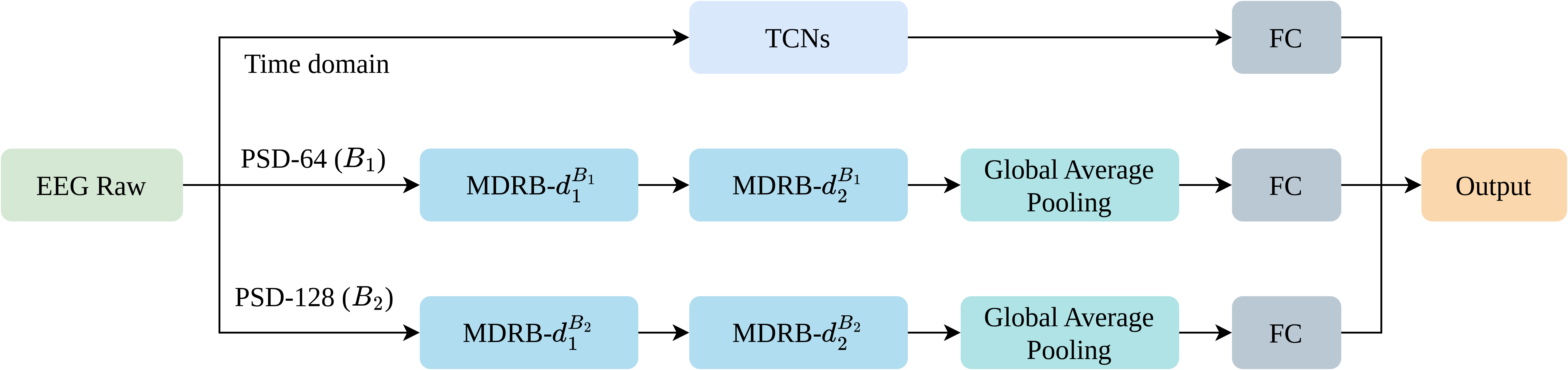}}
\caption{System overview: Combined time and frequency domains for mental workload estimation.}
\label{fig:residual_model}
\end{figure*}

\begin{figure*}
\centerline{\includegraphics[width=33pc]{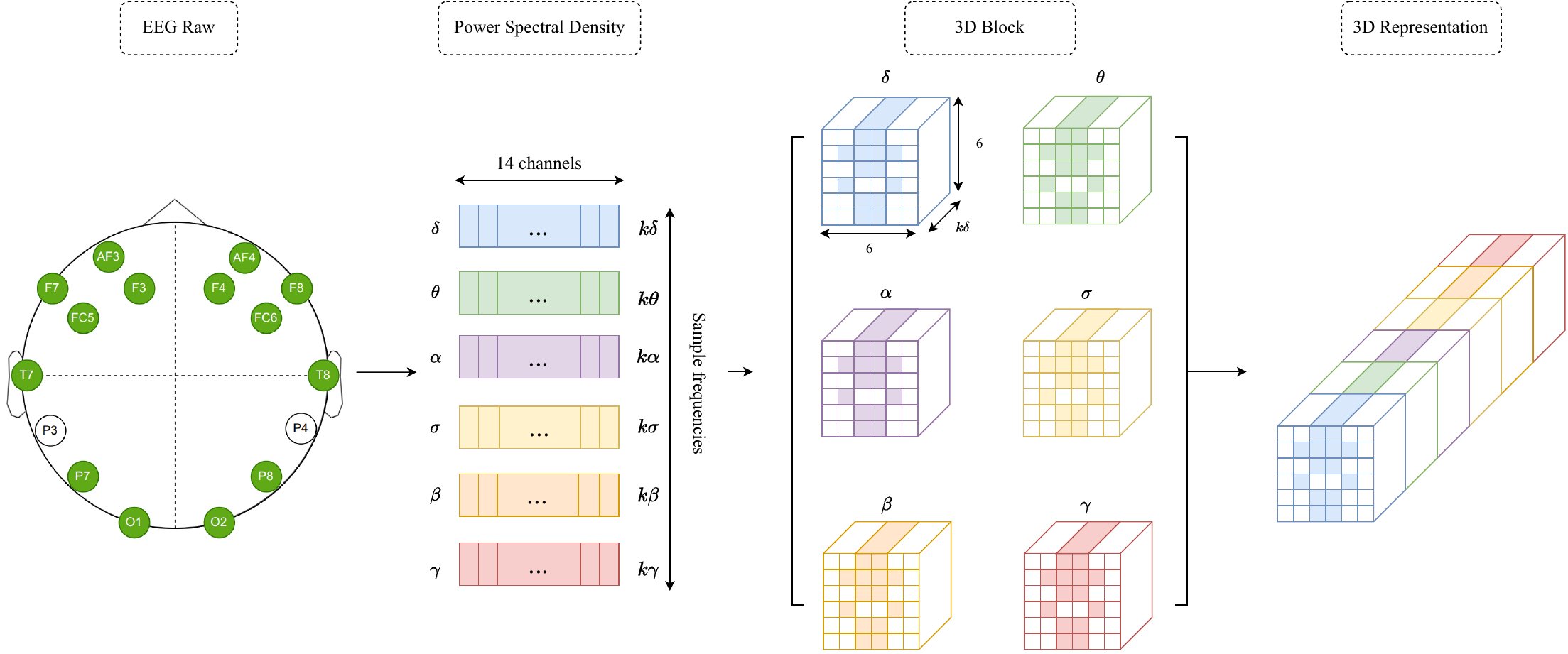}}
\caption{Overview of converting EEG signals to a 3D representation.}
\label{fig:psd}
\end{figure*}

\section{Proposed method}\label{sec:proposed_method}
\begin{figure}
\centerline{\includegraphics[width=15pc]{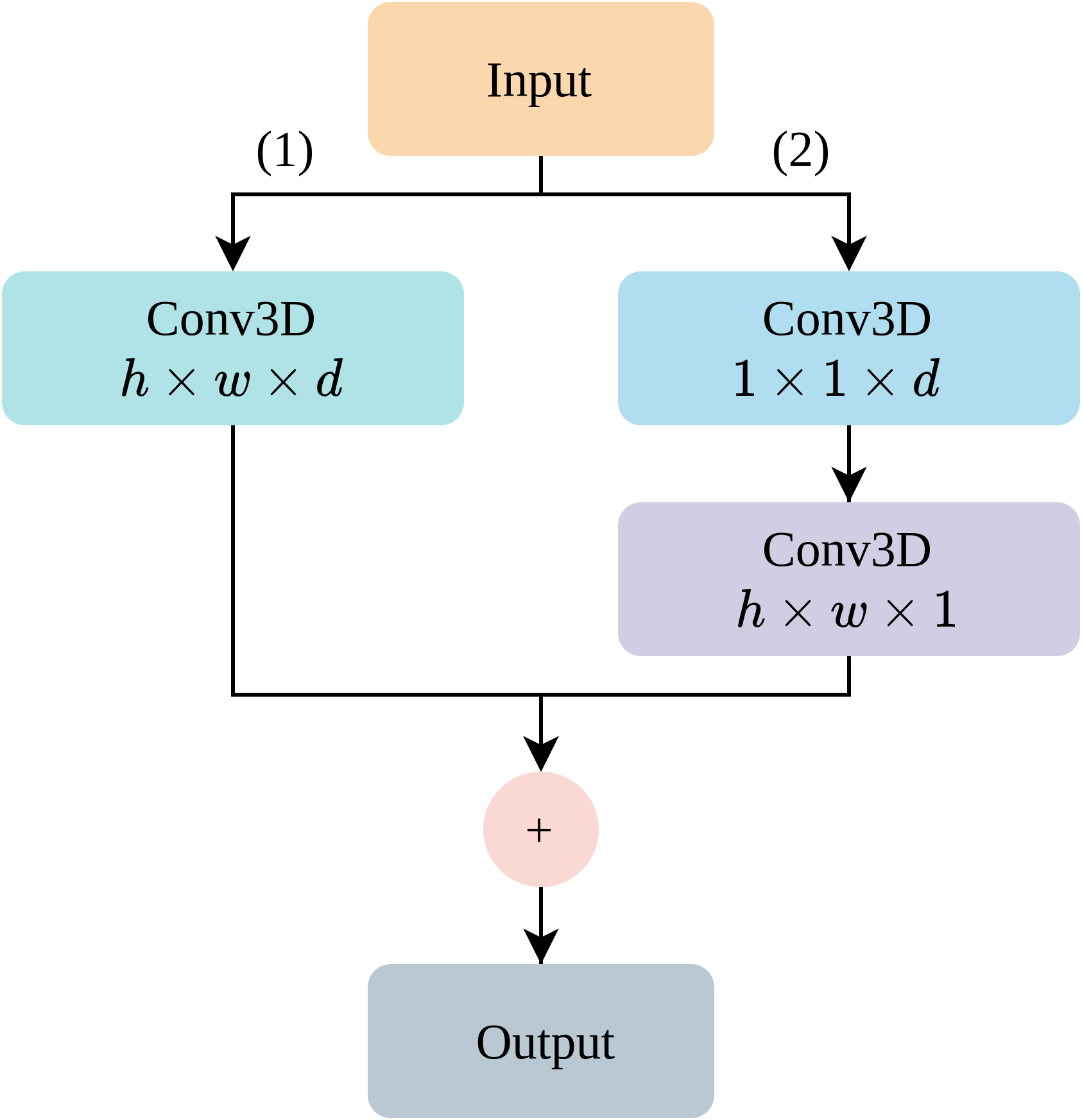}}
\caption{Multi-Dimensional Residual Block.}
\label{fig:residual_block}
\end{figure}

In this section, we present a deep learning model for estimating MWL based on both time and frequency domain information. We describe the 3D (three-dimensional) representation of PSD into 3D block as depicted in \autoref{fig:psd}. Additionally, we propose a new architecture, the Multi-Dimensional Residual Block (MDRB), illustrated in \autoref{fig:residual_block}. Finally, we provide an overview of our method in \autoref{fig:residual_model}.


\subsection{3D representation block}
We define $S^n = [E_1^n, E_2^n,...,E_W^n]^T \in \mathbb{R}^{W \times C}$ denote the $n^{th}$ sample, where $n \in {1,2,...N}$ and $N$ is the number of samples. Here, $W$ represents the window size, $C$ represents the number of channels in the EEG devices, and $E_i^n=[e_i^1, e_i^2,...,e_i^C] \in \mathbb{R}^C$ is the EEG signal at timestamp $i$, where $i \in {1,2,...,W}$. For analysis, we convert EEG signals from the time domain ($S^n$) to the frequency domain using Welch's method to obtain the Power Spectral Density (PSD). Each frequency band carries information about different brain activities and states. Therefore, frequencies link changes in brain activity with various stages of brain waves. Initially, Hans Berger identified two bands: alpha ($\alpha$) and beta ($\beta$). Subsequently, researchers introduced additional bands such as delta ($\delta$), theta ($\theta$), sigma ($\sigma$), and gamma ($\gamma$). In this study, we extract six frequency bands: delta (0.5-4 Hz), theta (4-8Hz), alpha (8-12Hz), sigma (12-16Hz), beta (16-30Hz), and gamma (30-45Hz). The extracted frequency bands depend on the window length of the Fast Fourier Transform (FFT). Therefore, the length of values in each band may differ, but the number of channels remains the same (14 channels). We define PSD for the six bands as follows:
\begin{equation}
    \centering
    \begin{split}
        \label{eqn:psd}
        P^n &= [P_{1\delta}^n, P_{2\delta}^n,...,P_{k\delta}^n,P_{1\theta}^n, P_{2\theta}^n,...,P_{k\theta}^n,P_{1\alpha}^n, P_{2\alpha}^n,...,P_{k\alpha}^n, \\
        &  P_{1\sigma}^n, P_{2\sigma}^n,...,P_{k\sigma}^n,P_{1\beta}^n, P_{2\beta}^n,...,P_{k\beta}^n,P_{1\gamma}^n, P_{2\gamma}^n,...,P_{k\gamma}^n]^T \in \mathbb{R}^{K \times C}
    \end{split}
\end{equation}

where $K=\sum_{i=1}^6 kb$ is the total length of frequency of the bands. $kb$ is the length of the frequency band $b$, $b \in \{\delta, \theta, \alpha, \sigma, \beta, \gamma\}$, $P_{ib}^n=[p_{ib}^1, p_{ib}^2,...,p_{ib}^C] \in \mathbb{R}^C$ is the signal frequency band $b$ at the frequency $i^{th}$.

\begin{figure} 
    \centering
    \begin{tabular}{c c}
         \includegraphics[width=0.31\columnwidth]{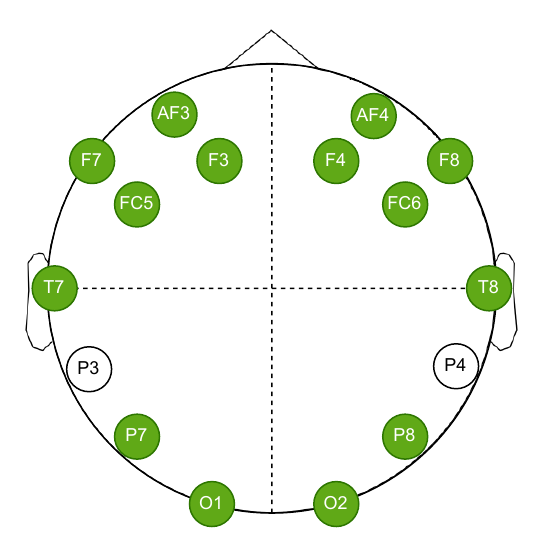} &
        \includegraphics[width=0.31\columnwidth]{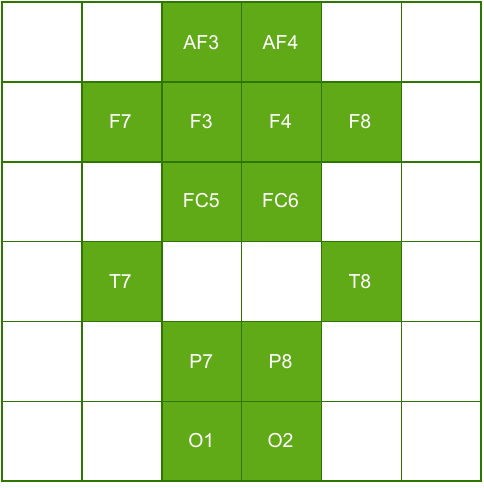} \\
        a) & b) \\
    \end{tabular}
   
\caption{Mapping of channels positions to a 2D matrix.}
\label{fig:convert_2d}
\end{figure}

Zhao et al. \cite{zhao2019multi} proposed a method for converting EEG data from 2D to 3D representation. The maximum number of horizontal and vertical electrodes is selected as the width and height of the matrix, respectively, based on the electrode positions. Fourteen signal positions ($P_{ib}^n$) are then mapped onto the 2D matrix with a size of $6\times6$. If an electrode position is empty, it is filled with a value of zero. The resulting 2D matrix is shown in \autoref{eqn:2d_matrix} and illustrated in ~\autoref{fig:convert_2d} b). 

\renewcommand{\arraystretch}{1.3}
\begin{equation}
\label{eqn:2d_matrix}
f(X_j) = 
\begin{bmatrix}
0 & 0 & AF3_j & AF4_j & 0 & 0 \\
0 & F7_j & F3_j & F4_j & F8_j & 0 \\
0 & 0 & FC5_j & FC6_j & 0 & 0 \\
0 & T7_j & 0 & 0 & T8_j & 0 \\
0 & 0 & P7_j & P8_j & 0 & 0 \\
0 & 0 & O1_j & O2_j & 0 & 0 \\
\end{bmatrix}
\end{equation}
where $X_j$ is the signal frequency of the band at the frequency $j^{th}$.

The extracted frequency bands step is shown in \autoref{fig:psd} as the Power Spectral Density step. From each band, we convert 2D matrix to 3D block with the size $6 \times 6 \times kb$. There are six blocks from six frequency bands (the 3D Block step as \autoref{fig:psd}). We stack them as a larger 3D block (the 3D representation step, illustrated in \autoref{fig:psd}). The height and width dimension keeps relationships between channels, and the depth dimension maintains information about the frequency and continuity of the bands. We define 3D block as $M^n=[M_1^n,M_2^n,...,M_K^n] \in \mathbb{R}^{6 \times 6 \times K}$, where $M_i^n=f(P_{ib}^n)$, $ib$ is the band $b$ at frequency $i^{th}$.


\subsection{MDRB: Multi-Dimensional Residual Block}
We propose a Multi-Dimensional Residual Block (MDRB) to capture the relationship between channels and frequency bands, as illustrated in \autoref{fig:residual_block}. The input to the MDRB is a 3D block of size $6 \times 6 \times K$, where $K$ is the depth of the 3D input. The MDRB has two branches. In the first branch, the input is passed through a 3D convolution with a filter size of $h \times w \times d$, where $h$, $w$, and $d$ are the height, width, and depth of the filter, respectively. The aim is to learn information about the relationships between channels and frequency simultaneously. In the second branch, the input passes through two 3D convolutions. The first convolution extracts information on the frequency bands using a filter size of $1 \times 1 \times d$. The second convolution extracts the relationships between channels using a filter size of $h \times w \times 1$. Finally, we compute the average of the outputs of both branches to obtain information from every observation of both branches.


\subsection{Combined time and frequency domain}
Our proposed model aims to combine the time and frequency domains, as depicted in \autoref{fig:residual_model}. To learn temporal information about the time domain, we utilize Temporal Convolutional Networks. For the frequency domain, we have two branches: PSD-64 ($B_1$) and PSD-128 ($B_2$), corresponding to two 3D blocks extracted from different lengths of FFT, computed by Welch's method. Each branch has a different depth, $d^{B_1}_{in}$ and $d^{B_2}_{in}$, respectively. We have PSD-64 ($B_1$) and PSD-128 ($B_2$) branches with the lengths of FFT 64 and 128, respectively. For each branch, the 3D block input passes 2 MDRBs, which have a different depth length of the MDRB, $d^i_1$, $d^i_2$, respectively, where $i$ is the PSD branch, $i \in \{B_1, B_2\}$. The value of MDRB depth (MDRB-$d^i_j, j \in \{1,2\}$) depends on the depth of the input block ($d^i_{in}$). The MDRB-$d^i_1$ has a depth $d^i_1$ which is calculated by getting the ceiling function of dividing by 2 as \autoref{eqn:ceil_func}, $d^i_2$ is calculated as \autoref{eqn:in_block_2}.

\begin{equation}
    \begin{split}
        \label{eqn:ceil_func}
        d^i_1 &= \lceil \frac{d^i_{in}}{2} \rceil
    \end{split}
\end{equation}

\begin{equation}
    \begin{split}
        \label{eqn:out_block_2}
        d^i_{out} &= \lfloor \frac{d^i_{in} + 2 \times pad - dil \times (d^i_1 -1) - 1}{str} + 1 \rfloor
    \end{split}
\end{equation}

\begin{equation}
    \begin{split}
        \label{eqn:in_block_2}
        d^i_2 &= d^i_{in} - d^i_{out}
    \end{split}
\end{equation}
where $pad$ is padding ($pad=0$), $dil$ is dilation ($dil=1$), and $str$ is convolution stride ($str=1$).

The Global Average Pooling was applied to the MDRB output for down sampling, and then the fully connected layer was attached to obtain the prediction. The mean of the predictions (3 branches) returns the prediction for the classification and estimation levels of MWL.


\section{Experiments}\label{sec:experiments}

\subsection{Dataset}
The STEW (Simultaneous Task EEG Workload) EEG database \cite{lim2018stew} includes 48 participants. Each participant needs to participate in two tasks, the `No' task and the SIMKAP (Simultaneous capacity) task. For the `No' task, participants do not perform any task and require a comfortable position with their eyes open. For the second task, participants need to solve the SIMKAP task, a test about a commercial psychological created by Schuhfried GmbH to evaluate a person's multitasking ability and stress. Both tasks were recorded for 3 minutes for the EEG signal. The first and last 15 seconds have been removed to eliminate the effects between tasks. A rating scale of 1-9 was used to assess MWL levels for each participant. 

The STEW dataset contains EEG data collected from the Emotiv headset. The device established 128Hz for sampling frequency, 16-bit A/D resolution, and fourteen channels (AF3, F7, F3, FC5, T7, P7, O1, O2, P8, T8, FC6, F4, F8, AF4) illustrated in ~\autoref{fig:convert_2d} (a) by 10-20 international system.

\subsection{Mental workload estimation}
We conducted experiments for classification and prediction of continuous levels. The classification task splits the rating scale into 3 classes of low (lo) workload (1-3 rating), middle (mi) workload (4-6 rating), and high (hi) workload (7-9 rating). The continuous level prediction task scales the MWL levels into the range [0-1] based on the rating scale from 1 to 9. The window length is 512 and the shift 128 for fourteen channels, which is also the parameter (window and shift size) that the STEW data used for their experiments. The dataset comprises forty-eight participants but ignores three because there are not available ratings for these participants. Therefore, 36 participants (80\%) were used for training and 9 (20\%) were used for testing.



\subsection{Experimental setup}
We used the TensorFlow framework to train and test our method. Our proposed method is trained in $150$ epochs and uses early stopping to obtain the best model. The Adam optimizer was employed with a learning rate of $0.001$ for classification and $0.0001$ for estimation of continuous levels. The TCNs \cite{KerasTCN} applied with the kernel size is $2$, the number of filters used for convolution layers is $128$, $2$ stacks of residual block were applied, and the dilation list is $[1, 2, 4, 8]$. The classification of three classes is a multi-label classification, so the Softmax function was used. The Sigmoid function was also utilized to get the output estimation for continuous levels of MWL.

In this study, categorical cross-entropy loss is used for classification. The loss function is calculated as follows:
\begin{equation}
    \begin{split}
        \label{eqn:cate_cross}
        CE &= -\sum_{i=1}^{N}y_i.log(\hat{y}_i)
    \end{split}
\end{equation}

And we used MSE (mean squared error) as a loss function for continuous levels estimation.
\begin{equation}
    \begin{split}
        \label{eqn:loss_function}
        MSE = \frac{1}{N}\sum_{i=1}^N(y_i - \hat{y}_i)^2
    \end{split}
\end{equation}
where $N$ is the number of samples, $y_i$ is the label of $i^{th}$ and $\hat{y}_i$ is the value prediction of $i^{th}$.

Additionally, we use the sensitivity, specificity, precision, negative predictive to evaluate the classification. The evaluations used are as follows:

\begin{equation}
   \begin{split}
       \label{eqn:sen}
       Sensitivity = \frac{TP}{TP+FN}
   \end{split}
\end{equation}

\begin{equation}
   \begin{split}
       \label{eqn:spec}
       Specificity = \frac{TN}{TN+FP}
   \end{split}
\end{equation}

\begin{equation}
   \begin{split}
       \label{eqn:prec}
       Precision = \frac{TP}{TP+FP}
   \end{split}
\end{equation}

\begin{equation}
   \begin{split}
       \label{eqn:neg}
       Negative\ predictive = \frac{TN}{TN+FN}
   \end{split}
\end{equation}



where TP is true positive, TN is true negative, FP is false positive and FN is false negative.

We used the Concordance Correlation Coefficient $\mathcal{CCC}$ function to evaluate for the estimation task. The $\mathcal{CCC}$ function is defined as:

\begin{equation}
    \begin{split}
        \label{eqn:ccc}
        \mathcal{CCC} &= \frac{2 \rho \sigma_{\hat Y} \sigma_{Y} }{\sigma_{\hat Y}^2 + \sigma_{Y}^2 + (\mu_{\hat Y} - \mu_{Y})^2}
    \end{split}
\end{equation}
where $\mu_{Y}$ is the mean of the label $Y$, $\mu_{\hat Y}$ is the mean of the prediction $\hat Y$, $\sigma_{\hat Y}$ and $\sigma_{Y}$ are the corresponding standard deviations,  $\rho$ is the Pearson correlation coefficient between $\hat Y$and $Y$.

\subsection{Results}
\autoref{tab:resutls} presents the results of our proposed method. We have split the dataset into an 8:2 ratio, following the methodology of Lim et al. \cite{lim2018stew}, to compare our results with their best single trial. The comparison shows that our proposed method outperforms Lim et al. \cite{lim2018stew} with an accuracy of 74.98\% compared to 69.0\%. We conducted separate and combined branch experiments, where the bold font defines the best performance. We also conducted seven experiments for each task that corresponds to the combination of 3 branches (Time, $B_1$, and $B_2$). It is noteworthy that combining of the branches gives better results than applying an individual branch. For the classification task, we obtained the best results $74.98\%$ for combining time and multi-frequency ($B_1$ and $B_2$) domain, which is better than the previous work \cite{lim2018stew} by extracting features of the delta, theta, alpha, and beta bands, $69.00\%$ ($28$ features from PSD) and $69.20\%$ ($56$ features from PSD) obtained, respectively. Furthermore, by combining the branch $B_1$ and branch $B_2$ without the time branch, we obtained 70.79\%, is better than Lim et al. \cite{lim2018stew}, 69.00\% respectively. It can be seen that the information from the frequency domain is helpful for MWL classification by combination.

\begin{figure}[!]
    \centering
    \begin{tabular}{c c}
    \includegraphics[width=0.5\columnwidth]{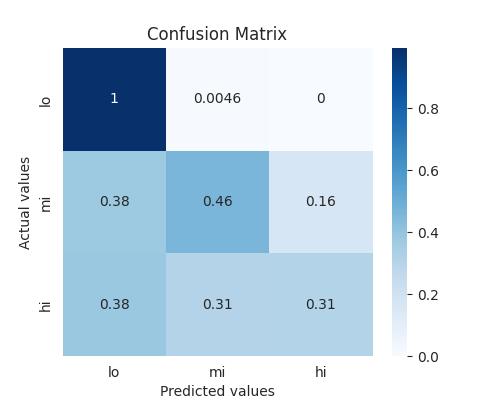}
    &
    \includegraphics[width=0.5\columnwidth]{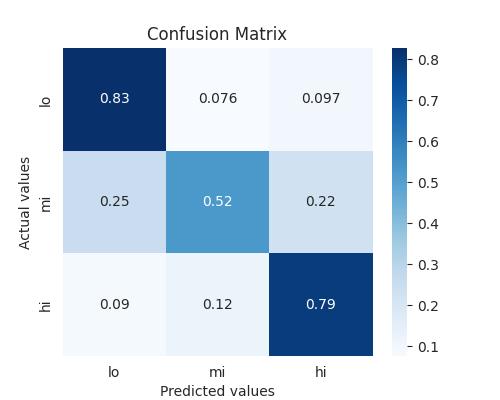} \\
    a) & b) \\
    \end{tabular}
    \caption{Confusion matrix for classification: (a) Lim et al. \cite{lim2018stew} with 28 features, (b) Our proposed method.}
    \label{fig:cfm}
\end{figure}

We used the Concordance Correlation Coefficient (CCC) metric to evaluate continuous estimation in our study. It should be noted that our use of CCC as a metric is relatively rare. Therefore, we recognize the need for further clarification and explanation of the results to enhance the reader's comprehension. CCC is a statistical measure widely employed in machine learning to evaluate the agreement between two measurement methods or observers for continuous-level estimation tasks, such as stress or valence/arousal estimation. Our proposed method achieved a CCC value of 0.629 for the continuous level estimation task. It is important to consider the specific context and application when interpreting the CCC values as high or low. A CCC value between 0.5 and 0.7 generally suggests moderate agreement, while a value above 0.7 indicates strong agreement. Therefore, the CCC value of 0.629 obtained in our study can be interpreted as demonstrating moderate agreement.

\autoref{fig:cfm} shows the confusion matrix of Lim et al. \cite{lim2018stew} (a) and our method (b). Figure (a) achieved the best for the 'lo' class, but the 'mi' and 'hi' classes are not good, as the only reached 0.46 and 0.31 respectively, which is less than 0.5 threshold. Our method improves for 'mi' and 'hi' classes, 0.52 and 0.79, respectively. In particular, our method obtained a significant result in the 'hi' class, which is 2.5 times greater than \cite{lim2018stew} ($0.31 < 0.79$). Generally, we have good results for classification, but our method is not really outstanding for the 'mi' class, with the result being much lower than the 'lo' and 'hi' classes. This shows our proposed focus on the 'lo' and 'hi' classes.






\renewcommand{\arraystretch}{1.5}
\begin{table}
    \centering
    \caption{MWL estimation: The results compare with previous work by accuracy and $CCC$ where $Acc_3$ is the classification for 3 classes.}
    \label{tab:resutls}
    \begin{tabular}{l l c c c}
    \\
        \hline
        & Feature & $Acc_{3} (\%) $ & $\mathcal{CCC}$ \\
        \hline
        Lim et al. \cite{lim2018stew} & 28 features from PSD & 69.00  & - \\
         & 56 features from PSD & 69.20 & - \\
        \hline
        Ours & Time & 56.92  & 0.341 \\
        & $B_1$ & 51.10 & 0.524 \\
        & $B_2$ & 62.89 & 0.510 \\
        & Time + $B_1$ & 34.32  & 0.609 \\
        & Time + $B_2$ & 36.28  & 0.477 \\
        & $B_1$ + $B_2$ & 70.79 & 0.525 \\
        & Time + $B_1$ + $B_2$ & \textbf{74.98} & \textbf{0.629} \\
        \hline
    \end{tabular}
\end{table}

\renewcommand{\arraystretch}{1.5}
\begin{table*}
    \centering
    \caption{The results for test set with change of shift on train set.}
    \label{tab:class_shift_resutls}
    \begin{tabular}{l c c c c c c c c}
    \\
        \hline
        Shift & \multicolumn{2}{c}{128} & \multicolumn{2}{c}{256} & \multicolumn{2}{c}{384} & \multicolumn{2}{c}{512} \\ \cmidrule(lr){2-3}\cmidrule(lr){4-5} \cmidrule(lr){6-7} \cmidrule(lr){8-9}
        & $\mathcal{CCC}$ & $Acc (\%)$ & $\mathcal{CCC}$ & $Acc (\%)$ & $\mathcal{CCC}$ & $Acc (\%)$ & $\mathcal{CCC}$ & $Acc (\%)$ \\
        \hline
        Time & 0.341 & 56.92 & 0.349 & 32.65 & 0.231 & 42.86  & 0.243 & 40.59 \\
        $B_1$ & 0.524 & 51.10 & 0.527 & \textbf{54.27} & 0.417  & 58.16  &  0.449 & 55.97  \\
        $B_2$ & 0.510 & 62.89 & 0.417 & 50.60 & 0.365 & \textbf{65.99} & 0.408 & 61.38 \\
        Time + $B_1$ & 0.609 & 34.32 &  \textbf{0.53}  & 27.78 & 0.302 & 27.70 & 0.310 & 27.44 \\
        Time + $B_2$ &  0.477 & 36.28 & 0.341 & 27.51 & 0.355 & 27.63 & \textbf{0.534} & 27.51 \\
        $B_1$ + $B_2$ & 0.525 & 70.79 & 0.303 & 51.25 & 0.285 & 50.30 & 0.352 & \textbf{62.43} \\
        Time + $B_1$ + $B_2$ &  \textbf{0.629} & \textbf{74.98} &  0.514 & 49.66 & \textbf{0.446} & 65.61 & 0.315 & 47.51 \\
        \hline
    \end{tabular}
\end{table*}



In our experimentation, we also varied the shift size when processing the training data. Specifically, we tried shift sizes of 128, 256, 384, and 512, which correspond to window overlaps of 75\%, 50\%, 25\%, and 0\%, respectively. The corresponding results are presented in \autoref{tab:class_shift_resutls}. Our findings indicate that a shift size of 128 results in better accuracy than the other shift sizes (256, 384, and 512).

\renewcommand{\arraystretch}{1.5}
\begin{table*}[]
    \centering
    \caption{Experiments of the MDRB if the branch $(1)$ or the branch $(2)$ is used.}
    \label{tab:run_branches}
    \begin{tabular}{c c c c c c c}
    \\
        \hline
        \multicolumn{1}{c}{} & \multicolumn{2}{c}{Only branch $(1)$} & \multicolumn{2}{c}{Only branch $(2)$} & \multicolumn{2}{c}{Combine branch} \\ \cmidrule(lr){2-3} \cmidrule(lr){4-5} \cmidrule(lr){6-7}
         Shift &  $Acc_{3} (\%) $ & $\mathcal{CCC}$ & $Acc_{3} (\%) $ & $\mathcal{CCC}$ &  $Acc_{3} (\%) $ & $\mathcal{CCC}$ \\
        \hline
         128 & 48.56 & 0.024 & 56.16 & \textbf{0.323} & \textbf{74.98} & \textbf{0.629} \\
         256 & 43.84 & 0.194 & 53.17 & 0.110 & 49.66 & 0.514 \\
         384 & \textbf{57.56} & \textbf{0.320} & 46.03 & 0.090 & 65.61 & 0.446 \\
         512 & 50.98 & 0.094 & \textbf{59.18} & 0.003 & 47.51 & 0.315 \\
        \hline
    \end{tabular}
\end{table*}


We conducted experiments to compare the performance of using only one of the two branches in the MDRB. The results are summarized in \autoref{tab:run_branches}. We observed that using both branches in combination yields the best result, with an accuracy of 74.98\% achieved with a shift value of 128. On the other hand, using only branch $(1)$ or branch $(2)$ results in lower accuracy, with 57.56\% and 59.18\%, respectively. For continuous labels, using an individual branch only achieves half the accuracy of the combined branches ($0.323 < 0.629$), indicating that combining information from both branches is effective. Branch $(1)$ captures an overview of information on both channels and frequency, while branch $(2)$ extracts detailed information in order of frequency and then channels.


\section{Discussion}\label{sec:discussion}
Our method utilizes both time and frequency domain information to extract valuable features that contribute to accurate classification and regression results. The experimental results demonstrate the effectiveness of combining these two domains, achieving good classification accuracy (74.09\%) and a concordance correlation coefficient ($CCC$) of 0.629 for regression. Our innovation lies in combining the time and frequency domains, which is an improvement over the previous work that used separate domain experiments. Additionally, our proposed method incorporates a Multi-dimensional Residual Block (MDRB) to extract information on channel relationships and frequency domain efficiently for the deep learning model. Furthermore, we conducted a novel experiment on the regression problem, which yielded good results and has the potential to contribute to the development of continuous-level mental workload state activity prediction systems.

Although we have achieved good results in MWL estimation, we have not thoroughly tested each channel or band. Our experiment involved using all EEG channels (14 channels) without performing specific experiments on each channel to evaluate their contributions to the MWL estimator process. In the frequency domain, we extracted information from six channels (delta, theta, alpha, sigma, beta, and gamma) and used them as a whole, but we did not perform separate experiments for each channel to assess the importance of each band in comparison to the effectiveness of combining them. These limitations need to be addressed in future research.

The STEW dataset has been widely used by researchers since its release in 2018. \autoref{tab:other_cpmpare_ref} provides the authors who used the STEW dataset for their study. However, there are variations in the segment inputs used by different authors, making it difficult to compare objectively such as Lim et al. \cite{lim2018stew} used a slide window of 512 with a shift of 128 as a baseline to perform experiments, the authors \cite{chakladar2020eeg, mohdiwale2020automated} used complete data duration, the authors \cite{zhu2021cognitive} used more than 3000 time points, and the authors \cite{taori2022cognitive} used 117 millisecond non-overlap.  In this study, we followed the input provided in the data, using a window size of 512 and a shift size of 128 for an objective comparison. We demonstrated that our method of classifying the three classes yielded higher results than the baseline ($74.98\% > 69.00\%$), and also achieved good accuracy in other assessments such as sensitivity, specificity, precision, negative predictive value (\autoref{tab:other_cpmpare_cfm}).

\renewcommand{\arraystretch}{1.5}
\begin{table*}[!]
    \centering
    \caption{Performance measures from the confusion matrix.}
    
    \label{tab:other_cpmpare_cfm}
    \begin{tabular}{l c c}
        \\
        \hline
            & Our proposed & Lim et al. \cite{lim2018stew}  \\
        \hline
        Sensitivity & \textbf{74.98} & 69.00  \\
        Specificity & \textbf{87.49} & 84.50  \\
        Precision & \textbf{74.98} & 69.00  \\
        Negative predictive  & \textbf{87.49} & 84.50  \\
        \hline
        
    \end{tabular}
\end{table*}

\renewcommand{\arraystretch}{1.5}
\begin{table*}[!]
    \centering
    \caption{Segment of previous work.}
    \label{tab:other_cpmpare_ref}
    \begin{tabular}{l c c c c}
        \\
        \hline
            & Year & Segment input & \multicolumn{2}{c}{Train-Test data $(\%)$} \\ \cmidrule(lr){4-5}
            & & & Train & Test \\
        \hline
        Lim et al. \cite{lim2018stew} & 2018 & Sliding window of 512 with shift of 128 & 80 & 20  \\
        Chakladar et al. \cite{chakladar2020eeg} & 2020 & Complete data duration & - & -  \\
        Mohdiwale et al. \cite{mohdiwale2020automated} & 2020 & Complete data duration & 80 & 20  \\
        Zhu et al. \cite{zhu2021cognitive} & 2021 & Greater than 3000 time points & 50 & 50  \\
        Taori et al. \cite{taori2022cognitive} & 2022 & Using 117 millisecond non-overlap & 80 & 20 \\
        \hline
    
    \end{tabular}
\end{table*}

\section{Conclusion}\label{sec:conclusion}
This paper presents our experiments on the STEW dataset, where we propose a method that addresses both the classification and the estimation of continuous levels of MWL. Our proposed method can be applied to other EEG data problems that involve labels, such as classification and estimation of continuous levels. Additionally, we introduce a Multi-Dimensional Residual Block (MDRB), which utilizes 3D residual blocks to learn channel relationships and information bands of the frequency domain. To improve the performance of the model, we combine the frequency and time domains. We use TCNs to learn temporal information for the time domain and combine it with two branches of the frequency domain. Our experimental results show that our proposed model outperforms the baseline of the STEW dataset. We also demonstrate that our multimodal fusion approach, based on multi-domain, leads to better results as each domain contributes differently to the MWL estimation process. However, we acknowledge the limitation of our model in classifying the middle stage of mental workload, which we aim to improve in the future. We plan to conduct experiments on other types of estimation, such as sleep stages, emotions, depression, etc., for EEG data.

\section*{Acknowledgments}
This work was supported by the National Research Foundation of Korea(NRF) grant funded by the Korea government(MSIT) (RS-2023-00219107). This work was also supported by Institute of Information \& communications Technology Planning \& Evaluation (IITP) under the Artificial Intelligence Convergence Innovation Human Resources Development (IITP-2023-RS-2023-00256629) grant funded by the Korea government(MSIT).

\section*{Declarations}

\begin{itemize}
\item Conflicts of interest: The authors declare no conflict of interest.
\item Authors' contributions: Methodology, H.-H.N. and S.-H.K.; Validation, H.-H.N. and S.-H.K.; Formal analysis, H.-H.N. and S.-H.K.; Investigation, H.-H.N. and S.-H.K.; Resources, H.-H.N. and S.-H.K.; Data curation, H.-H.N. and S.-H.K.; Writing—original draft preparation, H.-H.N.; Editing H.-H.N. N.K.I. and S.-H.K.; Review and editing, S.-W.K, H.-J.Y, S.-H.K.; Visualization, H.-H.N. and S.-H.K.; Supervision, S.-W.K, H.-J.Y, S.-H.K.; Project administration, S.-H.K.; Funding acquisition, S.-W.K, H.-J.Y, S.-H.K. All authors have read and agreed to the published version of the manuscript.
\item Ethical and informed consent for data used:  Not applicable.

\item Availability of data and materials: Publicly available at \url{https://ieee-dataport.org/open-access/stew-simultaneous-task-eeg-workload-dataset}

\end{itemize}

\bibliographystyle{unsrt}
\bibliography{sn-bibliography}

\end{document}